# Dependence of the onset of the runaway greenhouse effect on the latitudinal surface water distribution of Earth-like planets

T. Kodama[1,2], A. Nitta[1*], H. Genda[3], Y. Takao[1†], R. O'ishi[2], A. Abe-Ouchi[2], and Y. Abe[1]

[1] Department of Earth and Planetary Science, The University of Tokyo, 7-3-1, Hongo, Bunkyo, 113-0033, Tokyo, Japan

[2] Center for Earth Surface System Dynamics, Atmosphere and Ocean Research Institute, The University of Tokyo, 5-1-5, Kashiwanoha, Kashiwa, Chiba, 277-8568, Japan

[3] Earth-Life Science Institute, Tokyo Institute of Technology, 2-12-1 Ookayama, Meguro, Tokyo, 152-8551, Japan

Corresponding author: Takanori Kodama (koda@aori.u-tokyo.ac.jp)

(Current addresses)

[*]Tokyo Marine & Nichido Fire Insurance Co., Ltd.

[†]International Affairs Department, Tokyo Institute of Technology, 2-12-1, Ookayama, Meguro, Tokyo, 152-0033, Japan

**Key Points:**

- The onset of the runaway greenhouse effect depends strongly on surface water distribution.

- The runaway threshold increases as the surface water distribution retreats toward higher latitudes outside the Hadley circulation.

- The lower the water amount on a terrestrial planet, the longer the planet remains in habitable condition.

**Abstract**

Liquid water is one of the most important materials affecting the climate and habitability of a terrestrial planet. Liquid water vaporizes entirely when planets receive insolation above a certain critical value, which is called the runaway greenhouse threshold. This threshold forms the inner most limit of the habitable zone. Here, we investigate the effects of the distribution of surface water on the runaway greenhouse threshold for Earth-sized planets using a three-dimensional dynamic atmosphere model. We considered a 1-bar atmosphere whose composition is similar to the current Earth's atmosphere with a zonally uniform distribution of surface water. As previous studies have already showed, we also recognized two climate regimes: the land planet regime, which has dry low latitude and wet high latitude regions, and the aqua planet regime, which is globally wet. We showed that each regime is controlled by the width of the Hadley circulation, the amount of surface water, and the planetary topography. We found that the runaway greenhouse threshold varies continuously with the surface water distribution from about 130% (an aqua planet) to 180% (the extreme case of a land planet) of the present insolation at Earth's orbit. Our results indicate that the inner edge of the habitable zone is not a single sharp boundary, but a border whose location varies depending on planetary surface condition, such as the amount of surface water. Since land planets have wider habitable zones and less cloud cover, land planets would be good targets for future observations investigating planetary habitability.

**1 Introduction**

Liquid water is thought to be necessary for the emergence and evolution of life on the surface of planets [e.g., *Hart*, 1979; *Kasting et al.*, 1993; *Maruyama et al.*, 2013]. The conditions for the stability of surface liquid water on a planet have often been discussed in terms of the habitable zone (HZ). The HZ is defined as the region around the central star where liquid water is stable on the planetary surface. Throughout this paper, we call a planet with liquid water on its surface "a water planet", no matter how much liquid water exists.

When a planet with surface water receives insolation above a certain critical value, the planet is no longer able to maintain a thermal equilibrium state with surface liquid water. In this case, the excess of insolation is used to evaporate surface water until it is entirely vaporized [e.g., *Komabayashi*, 1967; *Ingersoll*, 1969; *Kasting*, 1988; *Abe and Matsui*, 1988]. A positive feedback of the strong greenhouse effect of water vapor causes the complete vaporization of liquid water in a runaway fashion. This effect is called the runaway greenhouse effect, and such a planetary condition is called the runaway greenhouse state. *Kasting* [1988] defined that the runaway greenhouse occurs when the critical point of water (~647 K and ~220 bar) is reached on a planet with an Earth-like water inventory. *Goldblatt et al.* [2013] and *Leconte et al.* [2013] defined that it occurs when there is an imbalance between the net absorbed solar flux and the outgoing infrared flux. The maximum insolation for which the planet can sustain a thermal equilibrium state with surface liquid water is called the runaway greenhouse threshold, which we simply call the runaway threshold. This threshold determines the location of the inner edge of the HZ.

There is another way to determine of the inner edge of the HZ, which is defined by the complete loss of the Earth's ocean over a timescale of 4.6 billion years [e.g., *Kasting et al.*, 1993; *Kopparapu et al.*, 2013]. When the insolation is small enough for the surface temperature to be below ~320 K, the mixing ratio of water vapor in the upper atmosphere ($f_{H_2O}$) remains quite low ($< 10^{-4}$) because of the cold trap. On the other hand, when the surface temperature is around 340 K due to high insolation, the cold trap moves upward and $f_{H_2O}$ suddenly increases. High $f_{H_2O}$ leads to rapid water loss through diffusion-limited hydrodynamic escape [*Hunten*, 1973; *Walker*, 1977]. When $f_{H_2O} = 3\times10^{-3}$, an amount of water comparable to the Earth's ocean can escape into space in 4.6 billion years, which corresponds to the age of the solar system. Such a planetary state is called the moist greenhouse state [*Kasting*, 1988; *Kasting et al.*, 1993].

When we consider the long-term evolution of the surface environment for terrestrial water planets, the onset of the moist greenhouse state is also important. This is because the onset of the moist greenhouse state would appear before the onset of the runaway greenhouse state, since main sequence stars—including our Sun—become brighter with time [e.g., *Gough*, 1981]. On the other hand, when we consider the temporal condition of planetary habitability, the onset of the runaway greenhouse state determines the location of the inner edge of the HZ.

Studies on the runaway greenhouse effect with a one-dimensional (1-D) radiative-convective equilibrium model of a gray atmosphere [e.g., *Nakajima et al.*, 1992] show that there is an upper limit of the infrared radiation from a planet with surface liquid water. This upper limit is called the Simpson-Nakajima limit [*Goldblatt et al.*, 2013]. In 1-D studies, the limit gives the runaway threshold [e.g., *Kasting et al.*, 1993; *Abe*, 1993; *Kopparapu et al.*, 2013; *Goldblatt et al.*, 2013]. The radiation limit for a fully water-saturated atmosphere is estimated to be 282 W/m$^2$ (102% $S_0$, where $S_0$ is the insolation at Earth's present orbit) [*Goldblatt et al.*, 2013] or 288 W/m$^2$ (104% $S_0$) [*Kopparapu et al.*, 2013].

Recently, the runaway threshold has been studied using general circulation models (GCMs) with three-dimensional (3-D) dynamics of the atmospheric circulation [*Abe et al.*, 2011; *Leconte et al.*, 2013; *Wolf and Toon*, 2014, 2015]. *Leconte et al.* [2013] has calculated the runaway threshold of the Earth by using the LMD (Laboratoire de Météorologie Dynamique) generic GCM. With the setups being those for the boundary conditions of the Earth at present, they found that the planet could preserve a thermal equilibrium phase with liquid water for an insolation of up to 110% $S_0$.

*Wolf and Toon* [2015] used the GCM, CAM (Community Atmosphere Model) ver.4, which was developed at the National Center for Atmospheric Research, and also investigated the stability of climate against the brightening Sun for the Earth at present. The purpose of their study was to clarify whether the moist greenhouse state would appear or not. They showed that the moist greenhouse state appears at 119% $S_0$. The climate is stable until 121% $S_0$. They also discussed the difference between their results and the results of *Leconte et al.* [2013], and pointed out the differences in treating moist physics, especially the formation of clouds.

These higher values of the runaway threshold from both studies are due to the effect of atmospheric circulation, especially in the drier regions, which is formed by the descending flow of the Hadley circulation. Moreover, the tops of the Hadley cells extend to higher altitudes and the descending flows dry the lower atmosphere [*Leconte et al.*, 2013]. The results obtained by both studies are qualitatively consistent with the results of a simplified GCM by *Ishiwatari et al.* [2002]. Thus, these studies indicate that the runaway threshold strongly depends on atmospheric circulation.

Whether aqua planets lapse into a moist greenhouse state has been investigated using 3-D GCM simulations [e.g., *Leconte et al.*, 2013; *Wolf and Toon*, 2015; *Popp et al.*, 2016]. *Leconte et al.* [2013] showed that the mixing ratio of water vapor in the upper atmosphere of Earth-like planets remains low until just before the onset of the runaway greenhouse effect. On the other hand, *Popp et al.* [2016] showed with 3-D aqua planet simulations using ECHAM6 and found that the transition from the cold regime, like present Earth's climate, to the moist greenhouse state driven by increasing solar forcing and $CO_2$ concentration. *Wolf and Toon* [2015] also found that the climate of Earth-like planets lapses into the moist greenhouse state before the onset of the runaway greenhouse effect. They also showed that if the incident solar radiation increases by about 2% of $S_0$ further from the incident solar radiation where the moist greenhouse effect occurs, the runaway greenhouse effect appears. They concluded that the complete water loss from a planet with Earth's ocean mass in the moist greenhouse state would occur before the onset of the runaway greenhouse state.

*Abe et al.* [2011] considered an idealized planet with the GCM, CCSR/NIES AGCM 5.4g, developed by the Center for Climate System Research, the University of Tokyo and the National Institute for Environmental Research for the study of the Earth's climate [*Numaguchi*, 1999]. They calculated the runaway threshold of a planet with a low amount of water on its surface—a "land planet"—and found that the runaway threshold is 170% $S_0$. This value is much larger than the runaway threshold for an aqua planet (135% $S_0$) estimated by the same GCM code in *Abe et al.* [2011]. It is much higher than that obtained by the 1-D vertical climate model or the 3-D GCM calculated by *Leconte et al.* [2013] and *Wolf and Toon* [2015]. On such a planet, when the obliquity is low, the surface water is localized around both poles and an extensive dry region appears at low latitude because the atmospheric circulation transports water vapor poleward [*Abe et al.*, 2005]. The planetary radiation in a low latitude region could be significantly higher than the Simpson-Nakajima limit estimated by 1-D calculation [e.g., *Nakajima et al.*, 1992; *Goldblatt et al.*, 2013] because the low latitude region is dry and the greenhouse effect of water vapor is very weak. Additionally, *Zsom et al.* [2013] focused on the relative humidity in the atmosphere and estimated the minimum distance for the inner edge of the habitable zone using a 1-D model. They also showed that the inner edge is located at 0.38 AU around a Sun-like star when the relative humidity is low and the surface albedo is high. Thus, these studies indicate that the runaway threshold depends on the surface water distribution as well as the atmospheric circulation.

The surface water distribution is determined by the balance between the water vapor transport due to atmospheric circulation and the liquid water transport on the surface. As shown schematically in Figure 1, atmospheric circulation transports surface water poleward due to the latitudinal temperature gradient, if the planet has a low obliquity and no surface water transport [*Abe et al.*, 2005, 2011]. Transported water vapor in cold regions condenses and precipitates to the surface at high latitudes. If the total amount of water on the planet is very low, precipitated water is trapped in a regional depression at high latitudes. Then, the distribution of surface water is determined by the relationship between precipitation and evaporation.

On the other hand, if the amount of water precipitated at high latitudes is larger than the capacity of the depression, it runs off and flows down to lower latitudes. This is surface water transport. The lowest latitude that the surface water flow reaches is determined by the water amount and topography, which are both intrinsic values for each planet. The flatter the topography, the lower the latitude the surface water can reach. With a steeper topography, a greater amount of water is needed to reach the lower latitudes. Hereafter, we define the lowest latitude that the surface water flow reaches as the "water flow limit (WFL)".

In this study, we classify water planets into three types: ocean planets, partial-ocean planets and land planets. An ocean planet is a planet that is fully covered by oceans and has no land on its surface [*Léger et al.*, 2004]. In this study, an ocean planet is sometimes described as a full-ocean planet to distinguish it from a partial-ocean planet. A partial-ocean planet has some land above sea level, but it has a certain amount of water on its surface at all latitudes [*Abbot et al.*, 2012]. This planet has a sufficient amount of water for its surface water transport to exceed its water vapor transport, and the surface water distribution is controlled by the surface water transport. The surface water transport is strong enough to keep the surface wet at all latitudes. The Earth is a partial-ocean planet. Therefore, both *Leconte et al.* [2013] and *Wolf and Toon* [2015] investigated the runaway thresholds for partial-ocean planets. In this study, partial-ocean planets and full-ocean planets are placed together in a group termed aqua planets [*Abe et al.*, 2011].

A planet with very little water localized around both poles is classified as a land planet [*Abe et al.*, 2005, 2011]. The degree of localization depends on the degree of surface water transport. The weaker the surface water transport, the more dominant the water vapor transport and the more localized the surface water distribution. Since *Abe et al.* [2011] completely removed surface water transport, the water is localized very close to the poles. Such a planet, which is what they considered, is an extreme case of a land planet.

Although there are large differences in the runaway threshold between an aqua planet and a land planet, no studies exist for the intermediate conditions; in other words, for the dependence of the runaway threshold on the distribution and transport of surface water. In this study, we clarify this dependence for Earth-like planets by performing numerical simulations with a GCM. Since the influence of 2-D surface water distribution and transport strongly depends on the planetary topography and bathymetry, we simplified these effects in our simulations by introducing a water flow limit (WFL) as a parameter. At higher latitudes than the WFL, the surface is always wet, but the surface conditions (wet or dry) are numerically predicted at lower latitudes than the WFL in the simulations.

In Section 2, we describe our method and introduce the concept of the WFL. In Section 3, we show some typical results and the dependence of the runaway threshold on the WFL. Then, in Section 4, we discuss the dividing factor between an aqua planet and a land planet, and the relation between the amount of water and the WFL for the topographies of Earth and Venus. In Section 5, we summarize our findings.

## 2 Methods

We performed a series of numerical experiments to clarify the dependence of the runaway threshold on the latitudinal surface water distribution. In this study, we calculated water vapor transport though the atmospheric circulation with vaporization and condensation of water by using a GCM, while we did not calculate surface water transport explicitly. Instead of this, we introduced a parameter called the "water flow limit (WFL)" in GCM calculations to describe the efficiency of the surface water transport. In GCM calculations, we assumed that a region at a higher latitude than this WFL is always wet, which means the water-saturated soil. On the other hand, the surface condition of a region at a lower latitude than this WFL is numerically calculated; these surface condition controlled by equatorward transport of water vapor. If this transport is efficient, the region at latitude lower than the WFL becomes wet. Here, we call the latitudinal boundary between the dry surface and wet surface the "dry edge" (see Figure 1).

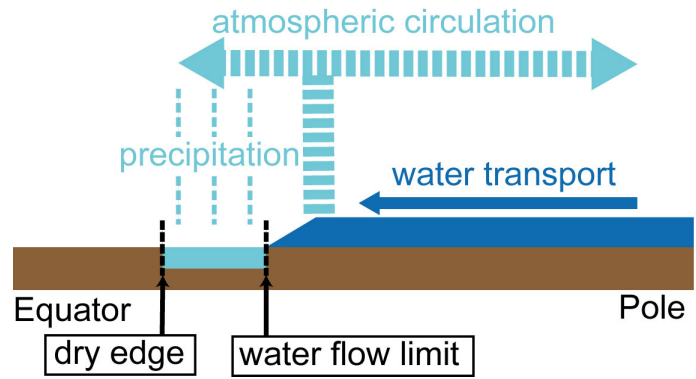

**Figure 1.** Schematic picture of the meridional water transport processes on the surface and in the atmosphere. The region at latitudes higher than the water flow limit is assumed to always be wet (see text for definition). On the other hand, the region at latitudes lower than the water flow limit is dry, if precipitation does not occur. If precipitation exceeds evaporation, the surface can become wet. The lowest latitude above which the surface is wet is defined as the dry edge.

The latitude of a numerically obtained dry edge should be lower than or equal to that of the WFL. Therefore, when the dry edge does not reach the equator, surface water is localized around both poles, and the wet regions are separated across the equator. This type of planet is a land planet. On the other hand, if the atmospheric water vapor transport is efficient enough to make the equator wet, the entire surface will be wet; that is, such a planet is an aqua planet. When the topography is given, the WFL is simply a function of the amount of water (see Section 4.2).

2.1 GCM description

We made an idealized water planet model by removing the ozone, topography and vegetation from CCSR/NIES AGCM 5.4g, which was developed for the climate modeling of Earth [*Numaguchi*, 1999] and was applied to the paleoclimate of Earth [*Abe-Ouchi et al.*, 2013]. For idealized planets, our setting corresponds to that in *Abe et al.* [2011]. We assumed that the planetary orbit is circular and the planetary obliquity is zero to avoid the effects of seasonal changes in the insolation.

The model processes are unmodified from their terrestrial use except for surface runoff, ocean horizontal heat flux, and the ozone radiation effect. Atmospheric composition, atmospheric surface pressure, planetary size (including surface gravity), and orbital period were unchanged from the current Earth's values.

The fundamental equations of dynamical processes are global primitive equations [*Haltiner and Williams*, 1980]. In the horizontal direction, the scheme of large-scale atmospheric circulation is the spectral transform method. The grid size is about 5.6° in longitude and latitude. In the vertical direction, we use the sigma coordinate in the grid discretization. The number of vertical layers is 20.

The scheme of cumulus precipitation is based on a simplified method of the Arakawa-Schubert scheme [*Arakawa and Schubert*, 1974; *Moorthi and Suarez*, 1992]. The scheme of a prognostic cloud water based on *Le Treut and Li* [1991] is applied for the large-scale condensation. Precipitation at the surface is determined to be snow or rain based on whether the wet bulb temperature at the lower atmospheric layer is below or above freezing point. The scheme of the radiative transfer in our model is based on the discrete ordinate method and the two-stream *k*-distribution method [*Nakajima and Tanaka*, 1986]. The radiative calculation is performed with 18 wavenumber channels that are divided into several subchannels. The total number of subchannels is 37. Both cumulus clouds and large-scale condensation are considered for the radiative calculation.

In all simulations, we ignored the transport of surface and underground water in GCM calculations. The effect of the surface water transport is expressed by retaining a wet surface at latitudes higher than the WFL. We used the bucket model [*Manabe*, 1969], where ground soil has the ability to contain liquid water up to the field capacity ($W_{g,max}$). When the soil moisture ($W_g$) is less than $W_{g,max}$, the change in $W_g$ is predicted as a net contribution of rainfall, evaporation and snow melt. If $W_g$ reaches $W_{g,max}$, the excess water is regarded as runoff. We set $W_{g,max}$ to be 5000 m in order to avoid runoff during the experiments. This value is sufficiently large because it is larger than the amount of water from precipitation.

In the GCM calculations, the soil moisture ($W_g$) is calculated in each grid. When $W_g$ is above 10 cm, we assumed that the surface is wet, and that the evaporation efficiency $\beta$ is unity. Conversely, we assumed the surface is dry for $W_g <$ 10 cm, and that $\beta$ is proportional to the soil moisture. Thus, the evaporation efficiency $\beta$ is expressed as

$$\beta = \min\{1, W_g/W_{g,\,crit}\},$$

where $W_{g,crit}$ is the critical value of soil moisture, which is set to be 10 cm in this paper. When $\beta$ = 1, the surface is completely wet. On the other hand, when $\beta$ = 0, the surface is completely dry. Using the calculated distribution of $\beta$, we classify an ocean planet as an aqua planet or a land planet. In this study, a planet with $\beta$ = 1 at its equator is defined as an aqua planet.

The planetary albedo is determined from the relationship between cloud and ground properties. Clouds are calculated in the GCM. In this study, the surface albedo ($\alpha$) is determined by whether or not the surface is covered with snow. A surface albedo with no snow ($\alpha_g$) is fixed at 0.3, which is the typical value for a desert, for simplicity. When the surface is covered with snow, the surface albedo over snow is assumed to be proportional to the square root of the thickness of the snow.

$$\alpha = \begin{cases} \alpha_g + (\alpha_s - \alpha_g)\sqrt{W_y/W_{yc}} & (W_y < W_{yc}) \\ \alpha_s & (W_{yc} \leq W_y) \end{cases},$$

where the mass of snow per unit area is $W_y$ and its critical value ($W_{yc}$) is 100 kg/m$^2$. Additionally, we consider the snow albedo ($\alpha_s$) with temperature dependence as follows,

$$\alpha_s = \begin{cases} \alpha_{sd} & (T_0 \leq T_d) \\ \alpha_{sd} - (\alpha_{sw} - \alpha_{sd})(T_0 - T_d)/(T_d - T_m) & (T_d < T_0 \leq T_m), \\ \alpha_{sw} & (T_m < T_0) \end{cases}$$

where $T_0$ is the surface temperature. The albedos of dry snow ($\alpha_{sd}$) and wet snow ($\alpha_{sw}$) are 0.7 and 0.5, respectively. Wet and dry snow represent whether melting occurs or not. The critical temperatures ($T_d$ and $T_m$) are 258.15 and 273.15 K, respectively.

An asymmetric surface water distribution is more realistic, as seen in the Earth today. However, to simplify, we assumed a symmetric water distribution between hemispheres in this study. For simplicity, we also assumed a zonally uniform distribution of surface water in GCM calculations. Hence, partial-ocean planets and (full-)ocean planets cannot be distinguished in this study.

2.2 Procedures for determining the runaway greenhouse threshold

The method to determine the dry edge and the runaway threshold for a given WFL consists of the following two steps. In the first step, the ground surface at latitudes higher than the given WFL is kept wet as the initial setting. We also put water of 10 cm depth on the surface at latitudes lower than the WFL as an initial condition. Since 10 cm of water on the surface makes the evaporation efficiency unity, the surface is globally wet at the initial setting. Then, assuming the present insolation of the Earth, we calculate water vapor transport in the atmosphere until a steady state is attained. The timescale for attaining steady states is about 10 years in our GCM calculations. In the calculations, the initially wet surface becomes dry when the evaporation exceeds the precipitation. On the other hand, the surface is kept wet where the precipitation exceeds evaporation.

In the second step, we typically increase the insolation by 1% $S_0$ (0.1% $S_0$ for land planets), and we perform another GCM calculation for 10 years for aqua planets and 20 years for land planets. In this calculation, we use the results obtained at the first step as the initial condition. We increased the insolation in a stepwise manner, and repeat GCM calculations by using the previous results as the initial condition until the climate never reaches a steady state. In practice, when the thermal equilibrium state breaks, the GCM calculation encounters an error and stops. In this study, the runaway threshold is defined as the upper limit of the insolation below which the planet can sustain thermal equilibrium with a certain amount of liquid water on the planetary surface, as in *Leconte et al.* [2013]. Figure 2 shows time-series plots of the bottom-of-atmosphere global-mean temperature and the top-of-atmosphere energy fluxes when the WFLs are 19.3° and 69.2°, respectively. When an imbalance between the incident solar flux and the outgoing planetary flux occurs, the surface temperature keeps increasing, which means that such a planet enters the runaway state.

Conducting these steps for various WFLs, we derived the relationship between the runaway threshold and the WFL. To clarify the boundary between aqua planets and land planets, we also derived the relationship between the WFL and the dry edge. The latitude of the dry edge is defined by the lowest latitude of a boundary between the wet surface and dry surface.

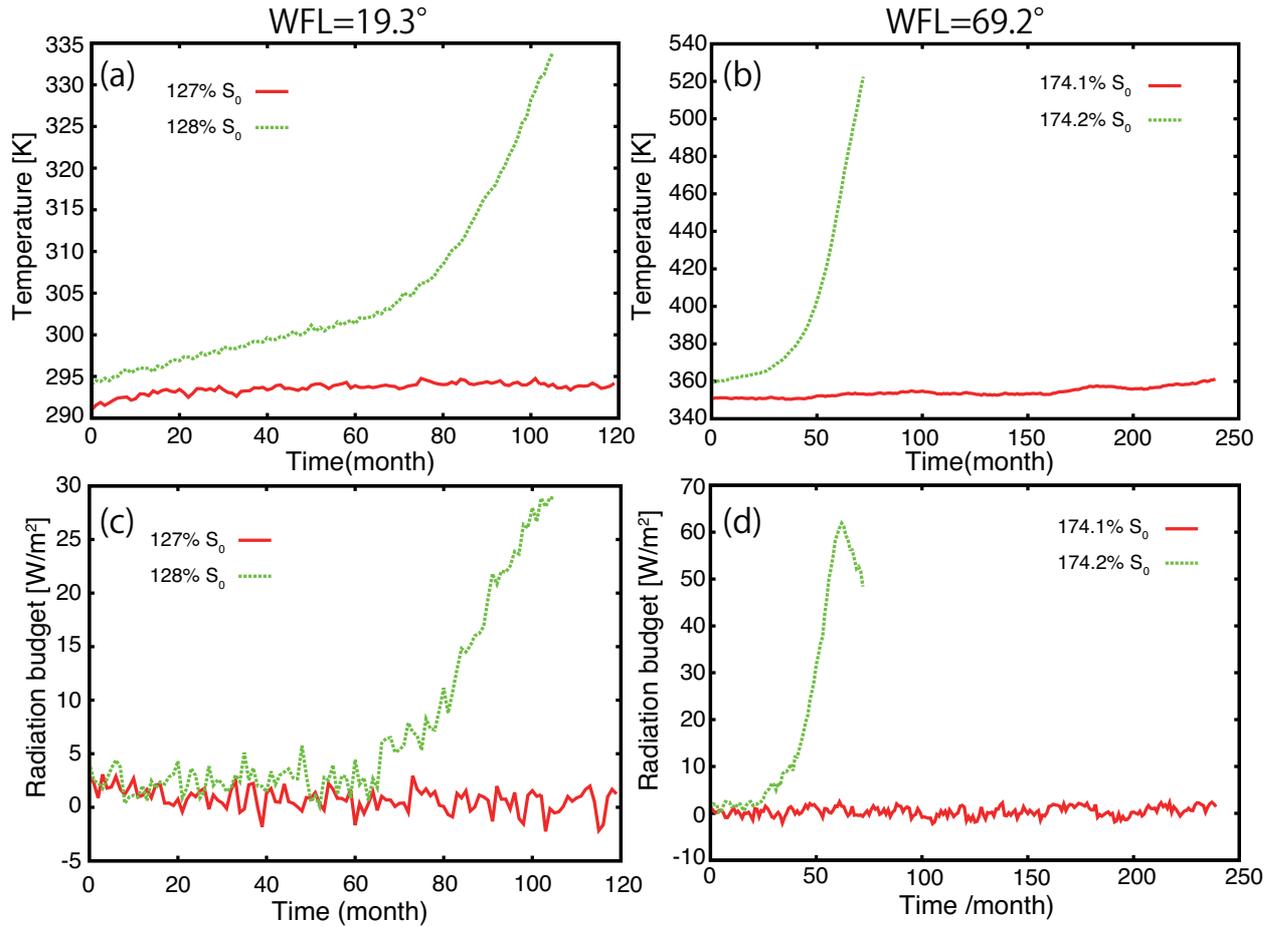

**Figure 2.** Time-series plot of the bottom-of-atmosphere global mean temperature (a and b) and the radiation budget at top-of-atmosphere (c and d) for a WFL of 19.3° and 69.2°, respectively. The red line and green dashed line show cases of a stable state and a runaway state, respectively. The present solar flux ($S_0$) is defined as 100%. The radiation budget is defined as the energy subtraction between the net solar flux and the outgoing planetary radiation. A month is 30 days.

## 3 Numerical Results

3.1 Dependence of the runaway threshold on the latitudinal surface water distribution

Figure 3 shows the dry edge latitude and the runaway threshold as a function of the latitude of the water flow limit (WFL). The runaway threshold is expressed as a percentage of $S_0$ (i.e., the present solar flux ($S_0$) is defined as 100%). We found that the runaway threshold varies continuously from 126% to 180% $S_0$ as the latitude of the WFL increases. *Abe et al.* [2011] have estimated the runaway threshold of a land planet to be 170% $S_0$ in the extreme case of a strongly localized surface water distribution, which is consistent with our result (170.8% $S_0$) for the WFL latitude 63.7°.

The results can be divided into two regimes: the land planet regime and the aqua planet regime. When the WFL is located at latitudes higher than about 25°, the calculated latitude of the dry edge is exactly the same as the given latitude of the WFL. Namely, the low latitude region stays dry. Since these planets are classified as land planets, we call this the land planet regime. In this regime, the runaway threshold strongly depends on the latitude of the WFL. This result indicates that the runaway threshold is higher when the transport of surface water is weaker.

On the other hand, when the WFL is located at latitudes lower than about 25°, which means that the WFL is located within the Hadley cells, the dry edge is located at the equator regardless of the WFL latitude. Namely, the surface stays wet globally. Since these planets are classified as aqua planets, we call this the aqua planet regime. In this regime, the runaway threshold is almost constant (~130% $S_0$) against the WFL latitude, which is almost comparable with the runaway threshold estimated by *Abe et al.* [2011]. If we reduce albedo on wet cells from 0.3 to 0.06 (ocean value), the runaway threshold on the aqua planet regime would to lower than the results from this study. However, there would be little change in the runaway threshold on the land planet regime because wet area on a land planet is narrow.

As a typical example of each regime at the runaway threshold, the results for the WFL located at 69.2° and 19.3° are shown in Figures 4 and 5, respectively. The calculated insolations of the runaway thresholds are 174.1% $S_0$ and 127% $S_0$ for the WFL latitudes of 69.2° and 19.3°, respectively. In the land planet regime, the atmosphere over latitudes lower than the WFL latitude (69.2°), which is the same as the latitude of the dry edge, becomes dry (Figure 4(c) and Figure 4(d)). From the mass stream function (Figure 4(e)), the direction of the atmospheric circulation in the lower atmosphere is poleward at high latitudes. As a consequence, the transport of water vapor in the atmosphere is only poleward, so that the dry edge is strictly located at the WFL. Such a dry atmosphere in the low latitude region can emit a large outgoing long-wave radiation (OLR) (Figure 4(b)). When the WFL and the dry edge are located at higher latitudes—in other words, the dry region on the planetary surface is broader—a land planet can emit a large OLR to space. As a result, a land planet can maintain liquid water on its surface at a higher insolation.

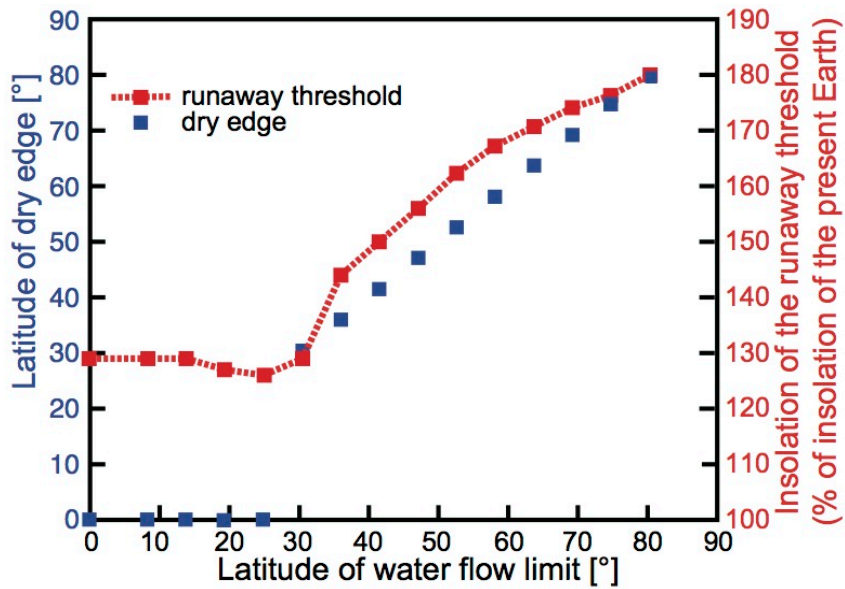

**Figure 3.** The dry edge latitude and the runaway threshold as a function of the water flow limit. The red and blue square symbols are the runaway threshold and the latitude of the dry edge, respectively. The latitude of the dry edge is equal to that of the WFL when the latitude of the WFL is larger than 30°.

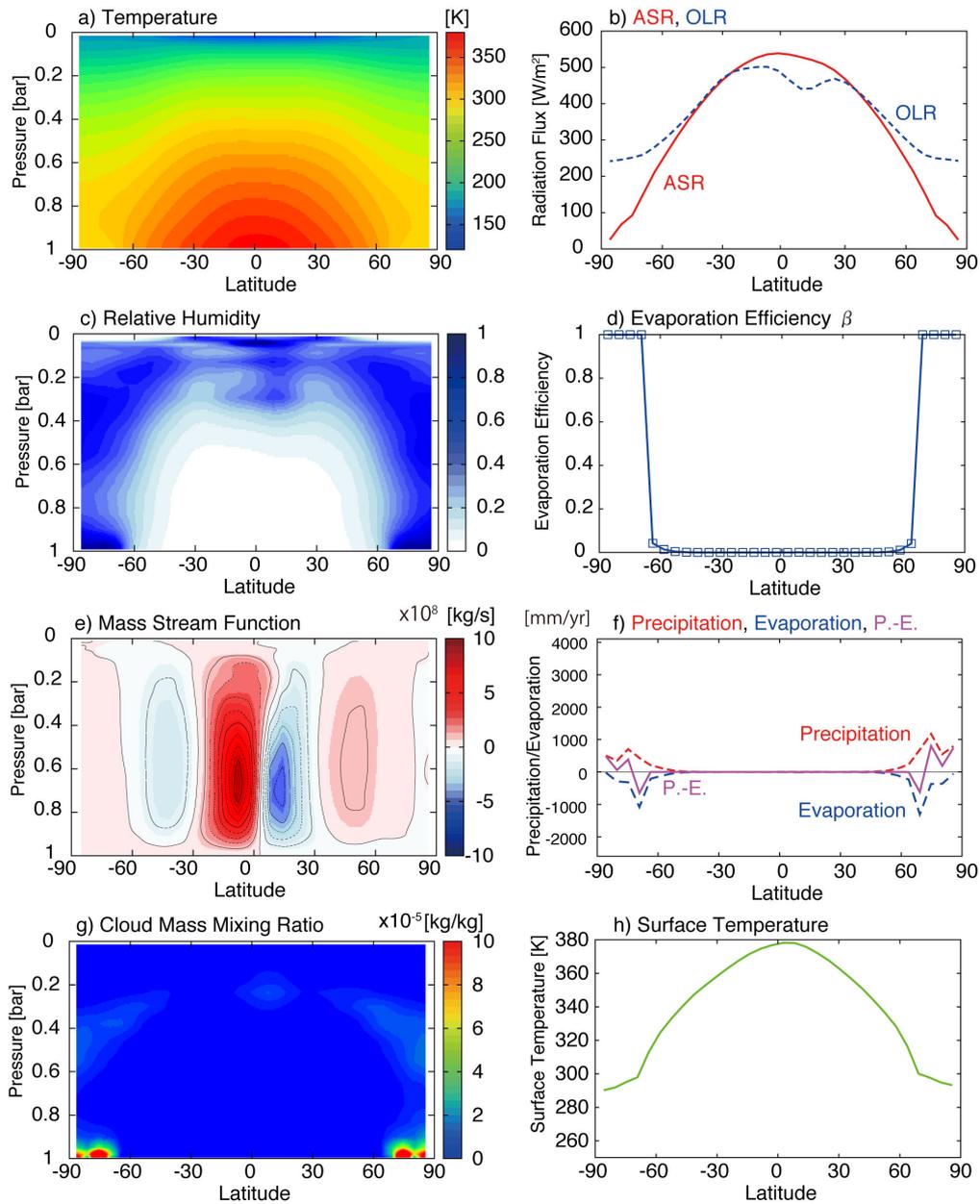

**Figure 4.** Climate variables for a typical land planet at the runaway threshold. Zonally averaged variables for the case of the water flow limit at 69.2° are shown. The region with latitudes lower than the water flow limit is found to be dry. Figures show that (a) the atmospheric temperature, (b) the absorbed solar radiation (ASR) and the outgoing long-wave radiation (OLR), (c) the relative humidity, (d) the evaporation efficiency *β*, which expresses ground wetness, (e) the mass stream function of meridional circulation, (f) the precipitation, evaporation, and difference between them (net precipitation), (g) the cloud mass mixing ratio and (h) the surface temperature.

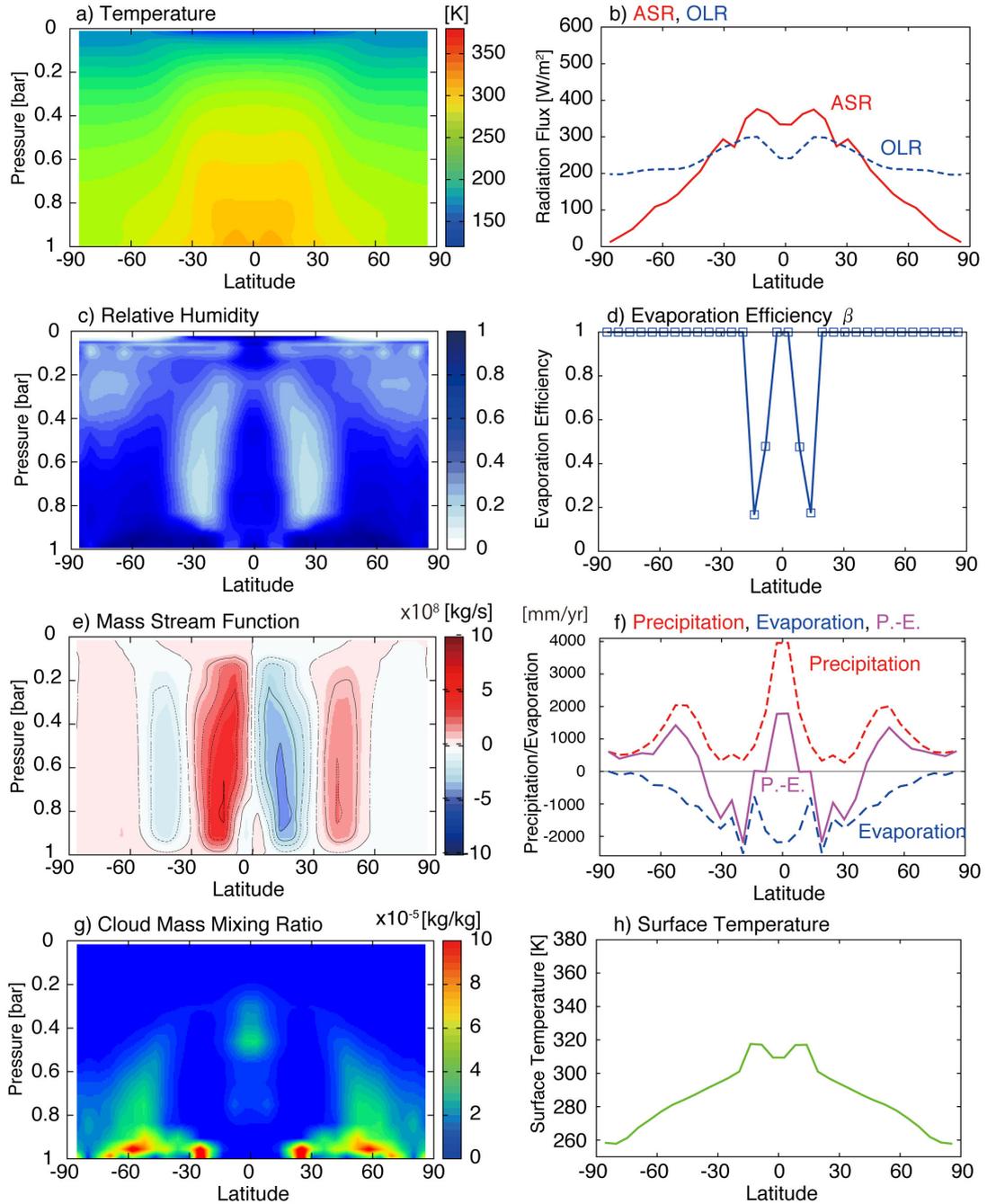

**Figure 5.** As in Figure 4, climate variables for a typical aqua planet at the runaway threshold. Zonally averaged variables for the case of the water flow limit at 19.3° are shown.

The climate in the aqua planet regime shown in Figure 5 is clearly different from that in the land planet regime. In this regime, the planetary atmosphere and surface become wet globally (see Figure 5(c) and (d)). Water vapor evaporated at the WFL is transported not only poleward, but also equatorward (see Figure 5(e)). Therefore, the atmosphere and surface are wet even at latitudes lower than the WFL. In this case, the dry edge is located at the equator; hence the entire surface is wet. The distribution of precipitation and evaporation is very similar to that observed on the Earth at present.

Such a wet atmosphere cannot emit an OLR greater than the radiation limit of the steam atmosphere (see Figure 5(b)). However, the OLR estimated by GCM calculations is greater than that obtained by 1-D simulations, as shown in *Leconte et al.* [2013] because there are regions not saturated with vapor that formed from the descending flow of the Hadley circulation in the planetary atmosphere. In this study, we found that the runaway thresholds for any WFL are almost constant (~130% $S_0$) in the aqua planet regime, as shown in Figure 3.

Figures 4(e) and 5(e) show the mass stream function of the meridional circulation. There is a Hadley circulation from the equator to about 25° of latitude in both hemispheres, and this latitude corresponds to the boundary latitude of the WFL that divides the aqua and land planet regimes. The Hadley circulation flows poleward in the upper troposphere, and equatorward near the surface. Therefore, if the latitude of the WFL is within the Hadley circulation flows (i.e., < ~25°), the lower portion of the Hadley circulation transports water vapor equatorward, which causes a globally wet atmosphere and ground.

In conclusion, the latitudinal relationship between the Hadley circulation and the WFL determines whether a water planet is a land planet or an aqua planet. If the WFL reaches the region of the Hadley circulations, the planet is in the aqua planet regime because the whole planetary surface becomes wet. The WFL is determined by the total amount of water on the planet and the topography, which will be discussed in Section 4.2. Since the Hadley circulation appears in the equatorial region, the width of the Hadley circulation is also important for this classification, which will be discussed in Section 4.1.

3.2 Characteristics of the onset of the runaway greenhouse state for the land planet regime

In this section, we examine characteristics of a land planet at the runaway threshold. Figure 6 shows the zonal mean OLR in the northern hemisphere at the runaway threshold for cases with various dry edge latitudes. Because the OLR at low latitudes is higher for a case with a dry edge at a higher latitude, the globally integrated OLR, namely the runaway threshold, becomes larger.

In some cases, the OLR has a depression in the low latitude region because of the greenhouse effect of clouds. The latitude of these depressions corresponds to the ascending region of the Hadley circulation. Under high insolation, the Hadley circulation changes its position slightly, northward or southward.

While the OLR from the dry low latitudes largely varies with the dry edge latitude, the OLR at the dry edge falls in the rather narrow range between about 250 and 300 W/m$^2$ (see Figure 6). For comparison, we evaluated the Simpson-Nakajima limit for the different relative humidities ($H_r$) of the atmosphere using the radiation code extracted from AGCM 5.4g (Figure 7). The radiation limit of this model depends on $H_r$. Since $H_r$ at the dry edge is about 20% to 60% in the GCM simulations as shown in Figure 4(c), the OLR at the dry edge at the runaway threshold (250–300 W/m$^2$) obtained in our GCM simulations (Figure 6) is similar to the Simpson-Nakajima limit (from 300 W/m$^2$ for $H_r$ = 0.6 to 380 W/m$^2$ for $H_r$ = 0.2 in Figure 7) estimated with 1-D models. This feature always appears when a land planet reaches the runaway threshold. Although the calculated values of the Simpson-Nakajima limit somewhat differ in the literature [*Goldblatt et al.* 2013; *Yang et al.* 2016], our result would qualitatively be unchanged even after applying the slightly different radiation code within an uncertainty.

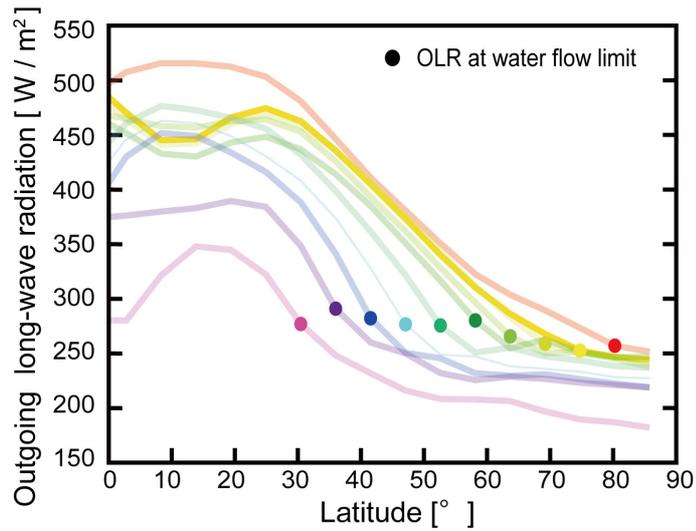

**Figure 6.** The zonal mean outgoing long-wave radiation (OLR) at the runaway threshold for various dry edge latitudes (corresponds to the latitude of the water flow limit). Each filled circle is the OLR at the water flow limit.

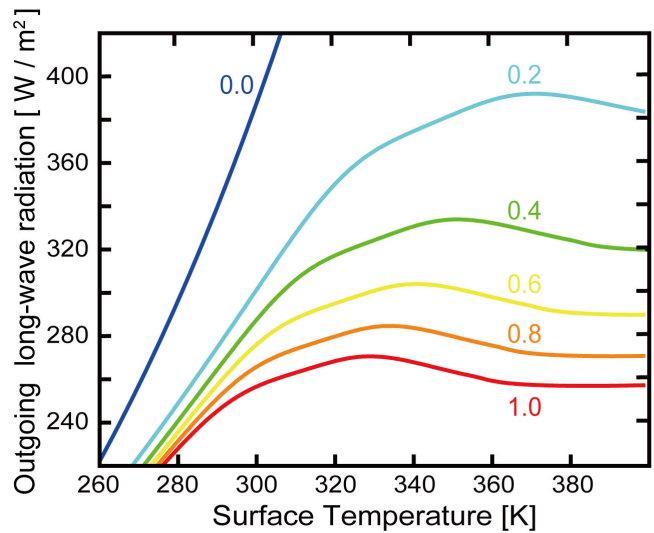

**Figure 7.** The outgoing long-wave radiation calculated for a 1-D atmosphere model with a radiation code used in AGCM 5.4g. An adiabatic troposphere with a given relative humidity and an isothermal stratosphere (150 K) are assumed. The values in this figure correspond to the relative humidity ($H_r$).

# 4 Discussion

4.1 The dividing factor between the aqua planet and land planet regimes

The strength of the atmospheric circulation is sensitive to various planetary parameters such as planetary rotation rate, planetary mass, surface gravity, insolation from the central star, the mass of the planetary atmosphere, obliquity, and so on. The width of Hadley circulation is important for classifying the climate of water planets as shown in the previous section. The width of Hadley circulation depends on various factors, such as the planetary rotation rate, the global mean temperature, the height of the Hadley cells, the equator-pole temperature difference and eddies [e.g., *Held and Hou*, 1980; *Satoh*, 1994; *Frierson* et al., 2007; *Kaspi and Showman*, 2015]. For instance, when the planetary rotation is faster, the region of the westerlies is wider. Thus, in such a case, the width of the Hadley circulation is narrower and the climate of such a water planet tends to take on the climate of the land planet regime. On the other hand, when the planetary rotation is slow, like Venus, the width of the Hadley circulation becomes wider and expands up to the polar region [e.g., *Pinto and Mitchell*, 2014] and such a water planet tends to adopt the climate of the aqua planet regime.

The results of this study show that neither the surface water distribution (the location of the dry edge) nor the insolation from the central star has much influence on the width of the Hadley circulation at the runaway threshold (see Figure 4(e) and 5(e)). Thus, the surface water distribution as well as the planetary rotation rate strongly affect the climate regime.

In this study, we considered planets with zero obliquity. When a planet has a non-zero obliquity, it develops seasonal climate change. Although the global annual mean insolation from the central star is hardly affected by the planetary obliquity, the planetary climate is affected by the planetary obliquity owing to the seasonal change. *Abe et al.* [2005] investigated the dependence of the climate's conditions on the planetary obliquity for a land planet using GCM. They found that planetary climates are divided into four climate regimes as functions of the summer surface temperature and the planetary obliquity: the warm-upright, warm-oblique, frozen-upright and frozen-oblique regimes.

When the summer surface temperature is above the freezing point of water (273 K), the planetary climates are classified into warm-upright or warm-oblique regimes. The major differences between these two regimes are the annual mean precipitation and the areas of rainfall. When the value of the obliquity is less than the value of the latitude for the width of the Hadley circulation, a land planet is in the warm-upright regime. On the other hand, when the value of the obliquity is larger than the value of the latitude for the width of the Hadley circulation, the climate of a land planet is in the warm-oblique regime. In the warm-upright regime, the low latitude area becomes dry and precipitation occurs at high latitudes owing to water transport in the planetary atmosphere. On the other hand, in the warm-oblique regime, precipitation occurs in the low latitude area in the summer hemisphere owing to the equatorward water transport because the temperature at the mid-latitudes is higher than that at the low latitudes in the summer hemisphere. Thus, the planet becomes wet globally. The boundary in obliquity between these two regimes is determined by the width of the Hadley circulation.

According to the results of *Abe et al.* [2005], the runaway threshold of a land planet obtained in our paper is applicable to the warm-upright regime. On the other hand, when the climate is in the oblique regime, the dry edge is located near the equator. Such a planet would behave as an aqua planet.

As mentioned above, the boundary in the WFL between the land planet regime and the aqua planet regime is determined by the width of the Hadley circulation. Also, the boundary between the upright and oblique regimes is determined by the width of the Hadley circulation. Thus, both climate boundaries are controlled by the planetary rotation rate. Although planetary rotation rates have not yet been measured for any terrestrial exoplanets (planets around other stars), future observations of their rotation rate and obliquities [*Kawahara*, 2012, 2016] are important for classifying the planetary climate through atmospheric dynamics.

In addition, we assumed circular orbits for the planets in our study; in other words, that the eccentricity of the planetary orbit is zero. A difference in the intensity of insolation between the pericenter and the apocenter appears in the case of non-zero eccentricity. When a planet has a large eccentricity, it is possible for it to receive a strong insolation above the radiation limit at the pericenter and to become cold below the freezing point of water at the apocenter. However, when a planet has a large heat capacity, such a large climatic change should not occur. Aqua planets have oceans, which have a large heat capacity. Thus, the effect of the eccentricity on the planetary climate for an aqua planet should be smaller than that for a land planet, which has a small heat capacity. Thus, when a planet has a large eccentricity, it is possible that an aqua planet may have a wider HZ than a land planet.

4.2 The relationship between the water flow limit and the water amount

The WFL is closely related to the total amount of water on the planetary surface and the planetary topography. In this section, we discuss those relationships using the current topography of Earth and Venus.

Using the current topographies of Earth, Venus and Mars [*Hirt et al.*, 2012; *Rappaport et al.*, 1999; *Wieczorek*, 2007], we calculated the WFL and the land fraction as a function of the amount of water (Figure 8). In these calculations, water was poured out from both poles; hence the water flowed out from high to low latitudes. The water was stored in depressions at higher latitudes than the WFL. Although the edge of the water flow depends on the longitudinal distribution of the topography, we assumed here that the lowest latitudinal edge of the water flow is the latitude of the WFL. This is because longitudinal water transport is faster than latitudinal water transport, and the longitudinal surface would become wet uniformly.

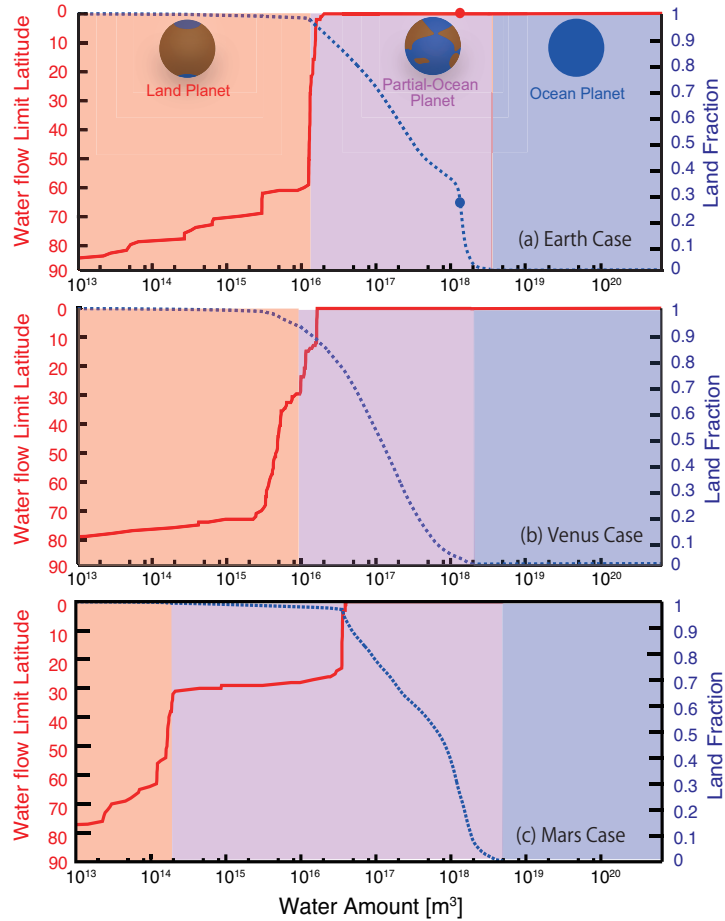

**Figure 8.** The latitude of the water flow limit (solid line) and the land fraction (dashed line) as a function of the amount of water for the topography of Earth (a), Venus (b) and Mars (c). In (a), red and blue filled circles are the latitude of the water flow limit and the land fraction for the present Earth, respectively.

When the amount of water is only the present volume of Earth's ocean ($V_{oce}$ = $1.35 \times 10^{18}$ m$^3$, e.g., *Menard and Smith* [1966]; *Genda* [2016]), the land fraction is 0.26 (c.f., 0.29 for Earth's present land fraction). When the amount of water reaches roughly $3V_{oce}$, the planetary surface is globally covered by the ocean (see Figure 8(a)). Such planets are (full-)ocean planets. Although the latitude of the WFL monotonically decreases with increasing amounts of water, it changes rather drastically beyond a certain amount. With the topography of Earth, a drastic change in the WFL occurs when the amount of water reaches $10^{16}$ m$^3$ (~ 1% of $V_{oce}$). This radical development of the WFL occurs when the amount of water corresponds to the capacity of the topographic lows around the North Pole (the Amerasian basin and the Eurasian basin) on the sea floor.

Provided that the width of the Hadley circulation is comparable to that of Earth at present, a planet with less than 1% of $V_{oce}$ should be in the land planet regime. When the amount of water is more than 1% of $V_{oce}$, the WFL reaches the Hadley circulation region and the planet belongs to the aqua planet regime, which is the case with the current Earth.

Caution must be paid to the fact that the present topography of Earth is affected by the present sea level—namely, the present ocean volume—through land erosion. Intrinsic factors (such as tectonics and mantle evolution) and external factors (such as land erosion) are responsible for the planetary topography. However, in spite of there being no ocean on Venus, the Venusian case shows a similar result to that of Earth (see Figure 8(b)). A drastic change in the WFL occurs around 1% of $V_{oce}$. Although the formation and evolution of planetary topography is important for planetary climate, it is beyond the scope of this study.

Even when the WFL reaches the equator, the land fraction is about 0.9 (see Figure 8(a)). This means that the area covered with water could be very small at the boundary between aqua and land planets, suggesting the importance of the longitudinal heterogeneity of water distribution for the climate of land planets. Planets with large ocean coverage should differ little from those fully covered with oceans. On the other hand, in the case of planets with a small ocean cover, it might be necessary to consider the longitudinal water distribution to estimate the runaway threshold for land planets with various amounts of water. However, it should be noted that the assumption of a zonally uniform distribution of surface water gives the wettest surface condition for a given WFL. Thus, the runaway threshold obtained from this study should be a lower estimate. We also assumed a symmetric water distribution between hemispheres. If a planet has an asymmetric water distribution between hemispheres, such as Mars does, we expect that the runaway threshold of such a planet can be determined by a lower latitude of the WFL. This is because even the ground at the equator should be wet when a water flow reaches the region of the Hadley circulation.

4.3 Evolution of water planets

Most studies on habitable planets so far have assumed Earth-like planets. Potentially habitable planets in extrasolar planetary systems are discussed on the basis of the knowledge gained from studies of Earth. The habitability of water planets is strongly affected by their atmospheric and climatic evolutions. Habitable planets with a wide variety of water amounts should exist in extrasolar systems because of the various formation processes of habitable planets.

Here, we discuss the long-term evolution of water planets in terms of the transition between the climate regimes shown above. Since main sequence stars become brighter with time, the insolation that water planets receive increases with time. Therefore, the climate of aqua planets is considered to enter the moist greenhouse state before the onset of the runaway greenhouse effect. In the moist greenhouse state, the upper atmosphere becomes moist. In this case, a rapid hydrogen escape via the photodissociation of $H_2O$ vapor occurs. When the mixing ratio of water vapor in the upper atmosphere is larger than $\sim 10^{-3}$, water comparable in total amount to Earth's oceans can escape into space in 4.6 billion years [e.g., *Kasting et al.*, 1993; *Kopparapu et al.*, 2013].

*Kodama et al.* [2015] systematically investigated whether the evolution from aqua planet to land planet occurs with different initial amounts of water and different orbital distances of the planet from the central star. They assumed the amount of water in the planetary atmosphere and on the planetary surface to be the transition conditions for the evolution from aqua planet to land planet.

They found that an aqua planet with a greater than 0.1 $V_{oce}$ evolves into the moist greenhouse state, followed by entering the runaway greenhouse state. This planet is no longer a water planet. On the other hand, an aqua planet with a less than 0.1 $V_{oce}$ can lose a significant amount of water into space without entering the runaway greenhouse state. Then the climate of this planet can evolve into the land planet regime. They suggested that a rapid water loss in the moist greenhouse state could extend the lifetime of habitability.

However, our study shows that the runaway thresholds of land planets gradually increase from the value of the runaway threshold of aqua planets (~130% $S_0$) to that of the extreme case for a land planet (~180% $S_0$) with increasing latitude of the WFL. Climatic evolution from the aqua planet regime to the land planet regime is not the only factor that can extend the lifetime of a habitable climate. Thus, whether land planets lapse into the moist greenhouse state or not is important to extending the lifespan of habitable conditions.

It is still unclear whether the climate of land planets would lapse into the moist greenhouse state or not. If the atmosphere of a land planet enters the moist greenhouse state, the surface water of the planet should decrease. At present, we are unable to estimate the amount of water in the upper atmosphere using AGCM 5.4g because the vertical water transport in the upper atmosphere is underestimated due to the vertical low-resolution in our simulation. Hereafter, we assume that water planets (aqua planets and land planets) can enter the moist greenhouse state, and discuss the evolution of water planets.

With the topography of Earth, the WFL changes rapidly around the water amount of 1% of $V_{oce}$, where the WFL changes from 25° to 58° (see Figure 8(a)). The typical timescale for insolation to increase by 2% of $S_0$ is of the order of 100 Myr, although it depends on the orbital distance of the planet. On the other hand, it takes as little as 10 Myr for a significant amount of water with 0.1% of $V_{oce}$ to escape through the diffusion-limited escape (the mixing ratio of water vapor being $10^{-3}$). The amount of water with 0.1% of $V_{oce}$ corresponds to the difference in the amount of water between the cases of the WFLs of 25° and 58°. The actual timescale of water escape should be shorter than 10 Myr because the mixing ratio of water vapor in the upper atmosphere increases with time during the period of the moist greenhouse state. Once an aqua planet enters the land planet regime, the WFL continuously moves poleward due to the decrease in the amount of water via water loss. If a land planet can enter the moist greenhouse state, it allows effective water loss. Thus, aqua planets with < 0.1 $V_{oce}$ survive for a longer time until all the water is lost or the insolation enters the runaway threshold of a land planet.

Focusing on the amount of water on the planetary surface, there are three evolutionary paths and two final fates of habitable planets (see Figure 9). As the central star evolves, an aqua planet with a greater than 0.1 $V_{oce}$ enters the moist greenhouse state, followed by the runaway greenhouse state. This evolutionary path was already proposed in previous studies [e.g., *Kasting et al.*, 1993].

An aqua planet with a less than 0.1 $V_{oce}$ can evolve from an aqua planet to a land planet through rapid water loss during the moist greenhouse stage. When the amount of water on the planetary surface reaches 0.01 $V_{oce}$—which corresponds to the boundary between the aqua planet regime and the land planet regime—the planet evolves into a land planet. If land planets enter the moist greenhouse state, they lose water on the planetary surface and the WFL moves poleward during the moist greenhouse state. The runaway threshold for a land planet becomes higher with an increase in the latitude of the WFL. Thus, a land planet can extend the duration of habitability with an increase in the latitude of the WFL (see Figure 9).

The third evolutionary path is the case for a planet that starts out as a land planet. Such a planet can maintain liquid water until it lapses into the runaway greenhouse state. As discussed above, it is possible for a land planet to evolve in such a way that it has a decreasing amount of water when it passes into the moist greenhouse state. This planet will finally enter the runaway greenhouse state.

Focusing on the amount of water on the planetary surface, there are three evolutionary paths discussed above, and two fates of habitable planets that ultimately end in the runaway greenhouse state, either from the aqua planet regime or the land planet regime (see Figure 9).

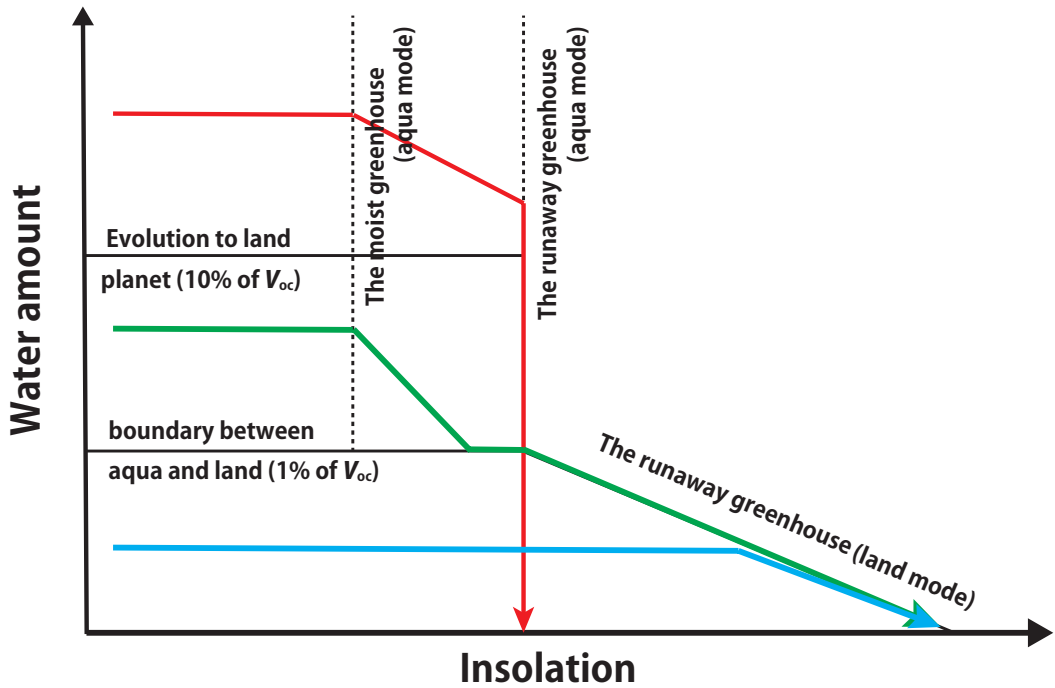

**Figure 9.** Schematic picture of evolution pathways for water planets. Aqua planets with a larger amount of water eventually lapse into the runaway greenhouse state (red line). On the other hand, aqua planets with less than 10% of $V_{oce}$ evolve into land planets, as discussed in section 4.3 (green line). If land planets enter the moist greenhouse state, such planets evolve in such a way that the amount of water they contain continues to decrease. For the case of a planet that starts out as a land planet, such a planet can maintain liquid water until it lapses into the runaway greenhouse state (blue line).

4.4 Implications for extrasolar planets

In the last two decades, the number of detected exoplanets has been increasing day by day, and some of them are Earth-sized or super-Earth-sized rocky planets. Recently, a super-Earth-sized planet located within the HZ around Proxima Centauri, which is the closest star to the Sun, was detected [*Anglada-Escudé et al.*, 2016], and its habitability has been discussed with GCM simulations [*Turbet et al.*, 2016]. TRAPPIST-1 has seven terrestrial planets and their climate is also discussed using GCM. Such tidally-locked planets around low mass stars can easily enter the land (or dry) planet regime because the night-side hemisphere becomes wet (or icy) and the day-side hemisphere becomes dry, which is very similar to a land planet. Tidally-locked planets can easily enter the land (or dry) planet regime if the atmosphere is not thick (i.e., if there is a weak energy transport from day-side to night-side). This is because water precipitates on the night-side hemisphere and would freeze there, and the day-side hemisphere becomes dry, which is very similar to the land planet that we examined in this study.

In the next decade, the number of detections of transiting exoplanets around neighboring stars will rapidly increase via observations by TESS (Transiting Exoplanets

Survey Satellite proposed by NASA) and PLATO (PLAnetary Transits and Oscillations of stars proposed by ESA). Both of them in combination with the radial velocity method will tell us more detailed information about exoplanets, such as the internal structure of super-Earths through observed mass-radius relations. After these missions, the JWST (James Webb Space Telescope) and the TMT (Thirty Meter Telescope) will focus on potentially habitable exoplanets and characterize them.

The surface conditions of exoplanets are important for their habitability because the HZs are strongly affected by them, as was shown above. Although bulk composition can be roughly estimated through the observed mass-radius relations (i.e., the bulk density of the planet), it will be difficult to identify small amounts of surface water even with future observations. Moreover, it is practically impossible to distinguish aqua planets from land planets from their bulk densities because the difference in the amount of water between an aqua planet and a land planet is $\sim 2.3 \times 10^{-4}$ wt% of the planetary mass, assuming the planetary mass and topography of Earth.

Recently, water vapor in the atmosphere of exoplanets was detected by the observations of transmittance spectroscopy of hot-Jupiters [e.g., *Madhusudhan et al*., 2014; *Cridland et al.*, 2016]. Although some exoplanets may have liquid water on their surfaces, the detection of water vapor in the atmosphere is not directly linked to the amount of water on the surface. On the other hand, a surface ratio of the land and ocean on an exoplanet would give us information about a small fraction of liquid water as shown in Figure 7. An observational strategy for obtaining the land-ocean ratio on exoplanets has been discussed [*Fujii et al.*, 2010], which would lead to us being able to distinguish between aqua planets and land planets.

Land planets have an advantage in the discussion of habitability. The HZ for a land planet is generally wider than that for aqua planets [*Abe et al.*, 2011]. Therefore, if exoplanets within the HZ for land planets are included in future targets for detailed observations, the number of observational targets for the characterization of habitable planets would drastically increase.

Land planets have yet another advantage. Even if an aqua planet is located in the HZ and has water vapor in its atmosphere, the issue with clouds remains. In the case of Earth, clouds are distributed globally and mask the information of surface properties. They also obscure the information of spectra from spectroscopy for characterizing exoplanets. By contrast, clouds on a land planet are located around the polar region (see Figure 4(g)) because of the localization of liquid water on the planetary surface. Thus, a land planet has an advantage not only when it comes to the detection of major atmospheric components, but also in detection of biomarkers ($O_2$, $O_3$, $CH_4$ and so on). If the distribution of clouds and surface water around the poles is detected by highly accurate direct imaging with TMT, we might be able to distinguish aqua planets from land planets. We will be able to get closer to finding extrasolar terrestrial habitable planets by considering land planets as observable targets.

According to Kodama et al. (2015), where the evolution from aqua planet to land planet was considered, land planets cultivate a habitable environment for a long time. If it takes a long time for life to merge and evolve, land planets are clearly better targets for the detection of biomarkers.

**5 Summary**

A previous study showed that surface water distribution is important for habitability [e.g., *Abe et al.*, 2011]. Depending on the surface water distribution, water planets can be divided into two categories: aqua planets and land planets. Surface water distribution is controlled by water transport both on the planetary surface and in the planetary atmosphere. Thus, we have introduced a parameter called the water flow limit (WFL) in this study. This parameter represents the efficiency of water transport on the planetary surface. Then, we investigated the effect of the surface water distribution on the onset of the runaway greenhouse effect using the GCM, AGCM 5.4g. We called the insolation for its onset the runaway threshold.

From the 3-D simulations, we have confirmed that there are two regimes for the runaway threshold with respect to the WFL; namely, the aqua planet regime and the land planet regime. When the WFL is located inside the region of the Hadley circulation (0–25°), the equator on the planetary surface is wet. In this case, planets behave as aqua planets and have a constant runaway threshold (about 130% $S_0$). On the other hand, when the WFL is located at high latitudes and outside the Hadley circulation region (> ~25°), the planetary surface and atmosphere lower than the WFL are dry and the runaway threshold increases with an increase in the latitude of the WFL (from 130% to 180% $S_0$).

The latitude of the WFL is a function of the amount of water, when the topography is given. To relate the WFL to the amount of water, we considered the topography of the current Earth and Venus, and estimated the amount of water for various values of the WFL. We found that the amount of water at the boundary between the aqua planet regime and the land planet regime is approximately 1% of Earth's present ocean volume for the case of both Earth's and Venus's topography.

In the next decade, several missions for the characterization of extra-terrestrial planets are planned. They will tell us more detailed information about exoplanets through the bulk density of exoplanets. However, it is difficult to get information about the surface environments of exoplanets. Land planets will be interesting targets for the discussion of potentially habitable planets because of their specific advantages: a long period in which a habitable environment exists, and the distribution of clouds. These characteristics will make a significant impact on further surveys of terrestrial habitable planets.

**Acknowledgments**

We thank our anonymous referees for their thorough review and constructive comments. We thank Prof. Masahiro Ikoma and Prof. Eiichi Tajika for their constructive discussion and helpful comments that improved our manuscript. This work was supported by MEXT KAKENHI Grant Numbers JP23103003, JP17H06104 and JP17H06457. This work was also partly supported GCOE program "From Earth to Earths" (MEXT). CCSR/NIES AGCM 5.4g used in this paper has been developed for the Earth's climate by the Center for Climate System Research, the University of Tokyo and the National Institute for Environmental Research (see *Numaguchi et al.* [1999] for more information on GCM). Datasets that create all of figure in this study are available on website (http://ccsr.aori.u-tokyo.ac.jp/~koda/). We used Gtool3-dcl5 by the GFD-DENNOU Club for analysis (https://www.gfd-dennou.org). To access all simulation data, please contact the corresponding author, Takanori Kodama (koda@aori.u-tokyo.ac.jp).